\documentclass[%
reprint,
superscriptaddress,
nofootinbib,
amsmath,amssymb,
aps,
pre,
notitlepage,
showkeys]{revtex4-2}
\usepackage[colorlinks=true, urlcolor=blue, anchorcolor=blue, citecolor=blue,filecolor=blue,linkcolor=blue,menucolor=blue]{hyperref}
\usepackage{url}
\usepackage{graphicx}
\usepackage{lipsum}
\usepackage[export]{adjustbox}
\usepackage{mathtools}
\usepackage[T1]{fontenc}
\usepackage{amsmath}
\usepackage{amssymb}
\usepackage{stmaryrd}
\usepackage[utf8]{inputenc}
\usepackage{rotating}
\usepackage{comment}
\usepackage{url}
\usepackage{bbm}
\usepackage{soul}
\usepackage[normalem]{ulem}
\usepackage{tabularx,booktabs}
\usepackage{stmaryrd}
\newcolumntype{Y}{>{\centering\arraybackslash}X}

\usepackage{hyperref}  
\usepackage{subfigure}
\usepackage{xr}
\usepackage{tcolorbox}
\usepackage{paralist}
\usepackage{booktabs}
\usepackage{xcolor}
\usepackage{pifont}

\begin{document}


\title{Branching under First-Passage Resetting}

\author{Aanjaneya Kumar}
\email{aanjaneya@santafe.edu}
\affiliation{Santa Fe Institute, 1399 Hyde Park Road, Santa Fe, NM 87501, USA}
\author{James Holehouse}
\email{jamesholehouse1@gmail.com}
\affiliation{Santa Fe Institute, 1399 Hyde Park Road, Santa Fe, NM 87501, USA}

\date{\today}

\begin{abstract}
Many biological processes, from cell division to viral lysis, are triggered when an internal stochastic variable reaches a threshold. Here we introduce \emph{Branching under First-Passage Resetting}, a general framework in which replication events arise endogenously from first-passage dynamics rather than from externally imposed lifetime clocks. We show that the resulting population dynamics obey an exact renewal equation linking single-trajectory first-passage statistics to the population growth rate. This mapping shows that, for fixed offspring number and fixed mean replication time, stochastic timing fluctuations necessarily enhance growth relative to a deterministic clock. When offspring yield depends on the first-passage time, however, fluctuations have non-trivial effects and expose a fundamental yield-delay trade-off: waiting longer can increase the number of descendants, but delays all future lineages. Our framework allows us to address this optimization problem analytically, and upon application to bacteriophage lysis, gives an optimal lysis time and growth rate consistent with empirical data.
\end{abstract}

\maketitle

\emph{Introduction.} Branching processes are a fundamental component of dynamical systems \cite{watson1875extinction,bellman1948theory,harris1948branching, harris1963theory,asmussen1983branching,athreya2012branching}, arising whenever individual trajectories ``\emph{branch}'' to give rise to multiple descendant trajectories. They underpin a wide range of phenomena, including nuclear reactions \cite{harris1960neutron,williams2013random,pazsit2007neutron,zoia2014clustering}, contagion processes \cite{gleeson2020branching,keating2022multitype,jamieson2020calculation,dumonteil2013spatial, jacob2010branching, kumar2020improved}, and cascading failures in complex systems \cite{dobson2004branching,kim2010approximating,sun2019power}. Branching processes have also proved valuable in elucidating the order and extreme-value statistics of strongly correlated random variables \cite{lindvall1976maximum,brunet2011branching,ramola2014universal,ramola2015spatial,ramola2015branching,majumdar2025extreme}. Owing to these broad applications, the theory of branching has attracted research attention from disparate disciplines. Yet, most theoretical treatments assume that branching events occur at times governed by an externally imposed clock, often independent of the underlying dynamics.

\begin{figure}
    \centering
    \includegraphics[width=1.0\linewidth]{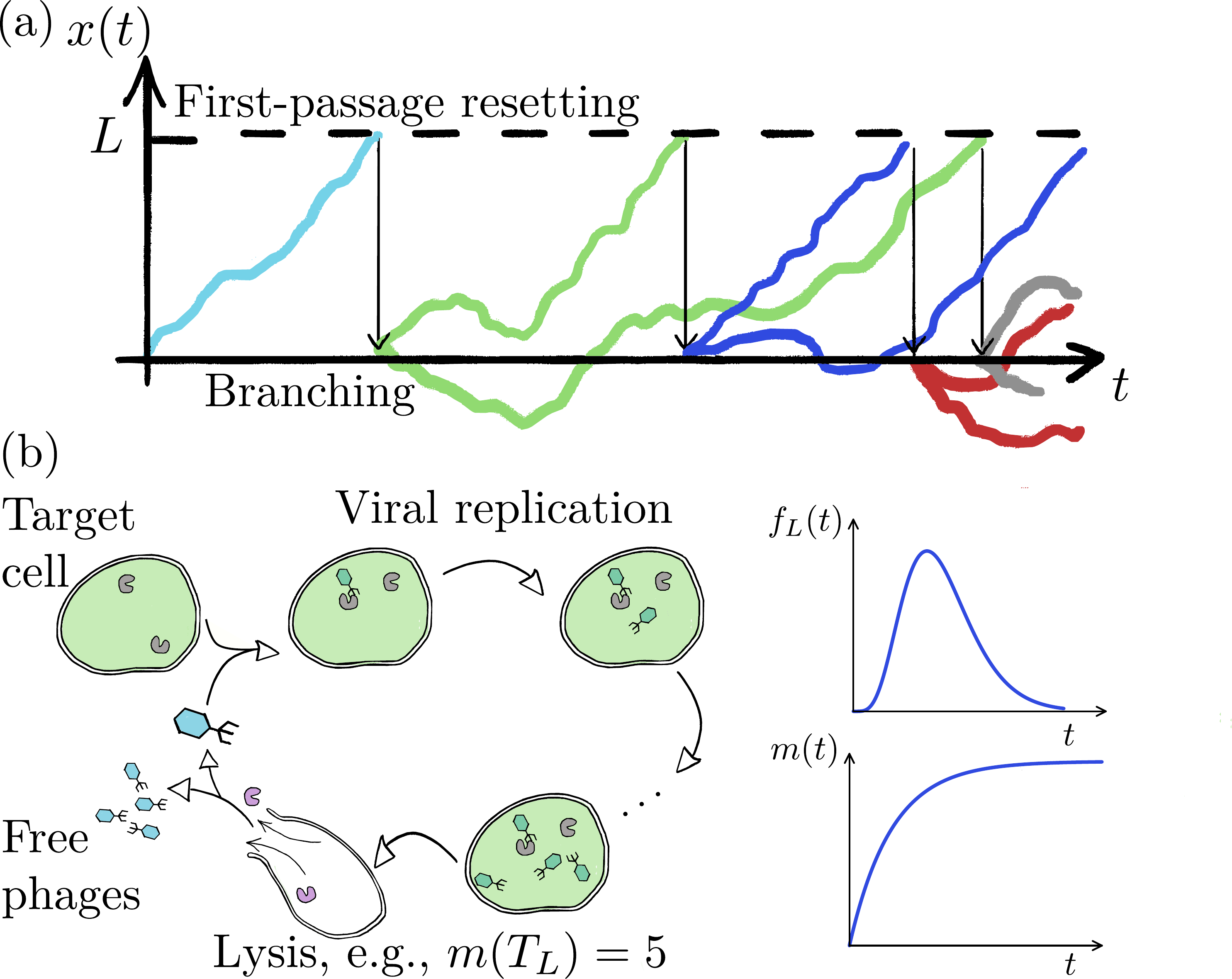}
    \caption{\emph{Branching under first-passage resetting.} (a) Schematic showing the core model for $m=2$. Trajectories stochastically evolve until the threshold is reached at which point the trajectory simultaneously resets to the origin and branches. (b) Processes such as viral lysis can be modeled under our framework, with distributions over the lysis time, $f_L(t)$, and a time-dependent number of offspring, $m(t)$.}
    \label{fig:placeholder}
\end{figure}

In stark contrast, branching events in many natural systems are generated \emph{endogenously}, triggered when an internal stochastic variable reaches a critical threshold. For example, bacterial cell division is fundamentally a branching event whose timing is set by the stochastic accumulation of growth-coupled division material to a threshold \cite{amir2014cell,si2019mechanistic}. Similarly, the timing of canonical bacteriophage lysis is governed by the concentration of membrane-associated holin protein reaching an effective triggering threshold \cite{wang2000holins}. In both scenarios, the timing of replication emerges organically as a first-passage event of an underlying stochastic process \cite{redner2001guide}. Figure~\ref{fig:placeholder} depicts a schematic illustration of this process. This viewpoint raises a natural question: \textit{how do single-trajectory first-passage statistics determine the emergent branching and population-level behavior?}

In this \emph{Letter}, we introduce \emph{Branching under First-Passage Resetting}, a general framework in which replication events occur when an underlying stochastic process crosses a prescribed threshold. At the threshold crossing instance, the original trajectory terminates and produces multiple independent offspring that restart from a common initial state. 
By employing a renewal approach, we derive an exact generalized Euler-Lotka equation that directly links microscopic first-passage statistics to macroscopic population growth. This allows us to derive a universal criterion demonstrating precisely when stochastic fluctuations in the underlying first-passage dynamics enhance the overall growth rate of the population. Furthermore, our framework exposes a trade-off fundamental to endogenous replication: when offspring yield is intrinsically coupled to the first-passage time, waiting longer to replicate can increase the potential burst size but penalizes the population growth rate by delaying subsequent replication. Our framework allows us to address this emergent optimization problem analytically. We finally specialize this optimization framework to bacteriophage lysis, yielding an optimal lysis time and growth rate in agreement with experimental measurements.

\emph{The Setting.} We consider a population of replicating entities whose proliferation is governed by an underlying stochastic accumulation process. The state of a single agent is described by a stochastic process $x(t)$ initialized at $x_0$ (chosen to be $0$ throughout). For brevity, we assume that $x(t)$ evolves in a continuous state space, but all of the analysis carried out here can be easily translated to discrete states. We define the first-passage time of this stochastic process, $T_L$, as the first time $x(t)$ crosses a predefined threshold $L$, or equivalently, 
\begin{equation}
T_L=\inf\{t>0:x(t) \geq L\}.
\end{equation}
We make no further assumptions on the dynamics of $x(t)$ beyond the existence of a probability density for $T_L$---a first-passage time density---which we denote as $f_L(t)$. Replication is triggered endogenously: when $x(t)$ first reaches the threshold $L$ at a random time $T_L$, the trajectory terminates and instantaneously produces $m$ identical offspring, each restarted at $x_0$ and evolving independently thereafter. The offspring number $m$ may be a fixed constant or a random variable, and its distribution may depend on the realized first-passage time.

Our central object of interest is the expected population density $\rho(x,t)$. Thus, the expected number of replicating entities whose internal state variable lies in $(x,x+dx)$ at time $t$ equals $\rho(x,t)\,dx$ and the expected population size is given by $N(t)=\int_{x<L} \rho(x,t)\,dx$. Note that $\rho(x,t)$ is a number density and not a normalized probability distribution unless $m=1$. Our focus on $\rho(x,t)$ is in contrast to much of the resetting literature, where the quantity of interest is the first-passage time \cite{Evans2020,evans2011diffusion,evans2011diffusion2,pal2015diffusion,pal2016diffusion,reuveni_optimal_2016,bhat2016stochastic,pal2017first,chechkin2018random}. Even in the cases where restart and branching have been considered together, the focus has been towards understanding how branching can expedite search times \cite{eliazar2017branching,pal2019first}, or studying the stochastic thermodynamics of such processes \cite{garcia2019linking,genthon2020fluctuation,genthon2022branching}, rather than understanding how branching and resetting affect the population's distribution over states.

\emph{Fixed yield.} We first consider the case where $m$ is a fixed constant. The dynamics of $\rho(x,t)$ are governed by the following renewal equation 
\begin{equation}
    \rho(x,t) = \rho_0(x,t) + m \int_0^t f_L(t') \rho(x, t-t') dt'
\label{eq:renewal}
\end{equation}
where $\rho_0(x,t)$ is the probability density of the initial particle having a state variable $x$ at time $t$ without having crossed $L$ in the meantime. The first term is therefore the no-replication contribution of the initial agent, whereas the second recursively sums the expected contributions of all sublineages initiated by first-passage events until time $t$.

Crucially, in the special case $m=1$, branching is suppressed and Eq.~\eqref{eq:renewal} reduces to a single trajectory undergoing renewal upon first-passage to the threshold. This is precisely the limit termed \emph{first-passage resetting} in Ref.~\cite{de2020optimization}, although related constructions appeared earlier \cite{sherman1958limiting,falcao2017interacting}. In recent years, first-passage resetting has found applications in modeling domain growth \cite{de2021optimization,de2022first}, wealth redistribution \cite{de2021tale}, dynamically emergent correlations \cite{biroli2026first}, and optimal target search \cite{biswas2025target,biswas2026optimal}. On the other hand, for general $m>1$, Eq.~\eqref{eq:renewal} takes the form of a generalized Bellman-Harris equation \cite{bellman1948theory}. The critical distinction here is that the branching-time distribution is not specified \emph{a priori}, but is generated entirely endogenously by the underlying first-passage statistics.

The Laplace transform of Eq.~\eqref{eq:renewal} can be rearranged to give
\begin{equation}\label{eq:laplacesoln}
    \tilde{\rho}(x,s) = \frac{\tilde{\rho}_0(x,s) }{1 - m\tilde{f}_L(s)},
\end{equation}
where the Laplace transform of function $g(t)$ is defined as $\tilde{g}(s) = \int_0^\infty e^{-st}g(t) dt$. In the long-time limit, the expected population size grows exponentially, where the growth rate $\lambda$ is determined by 
\begin{equation}
     m \tilde{f}_L(\lambda) \equiv m \langle e^{-\lambda T_L} \rangle = 1,
    \label{eq:master}
\end{equation}
which is akin to the celebrated \emph{Euler-Lotka equation}. It creates a direct map between the underlying first-passage dynamics and the population growth rate $\lambda$. Moreover, since $e^{-\lambda t}$ is strictly convex in $t$, applying Jensen's inequality yields,
\begin{equation}\label{jensen}
    \lambda \ge \frac{\ln m}{\langle T_L \rangle},
\end{equation}
establishing a rigorous lower bound on the population growth rate. This inequality dictates that stochastic fluctuations in the first-passage time will invariably enhance the macroscopic growth rate compared to a purely deterministic clock with the exact same mean division time $\langle T_L \rangle$.

\begin{figure}
    \centering
    \includegraphics[width=0.85\columnwidth]{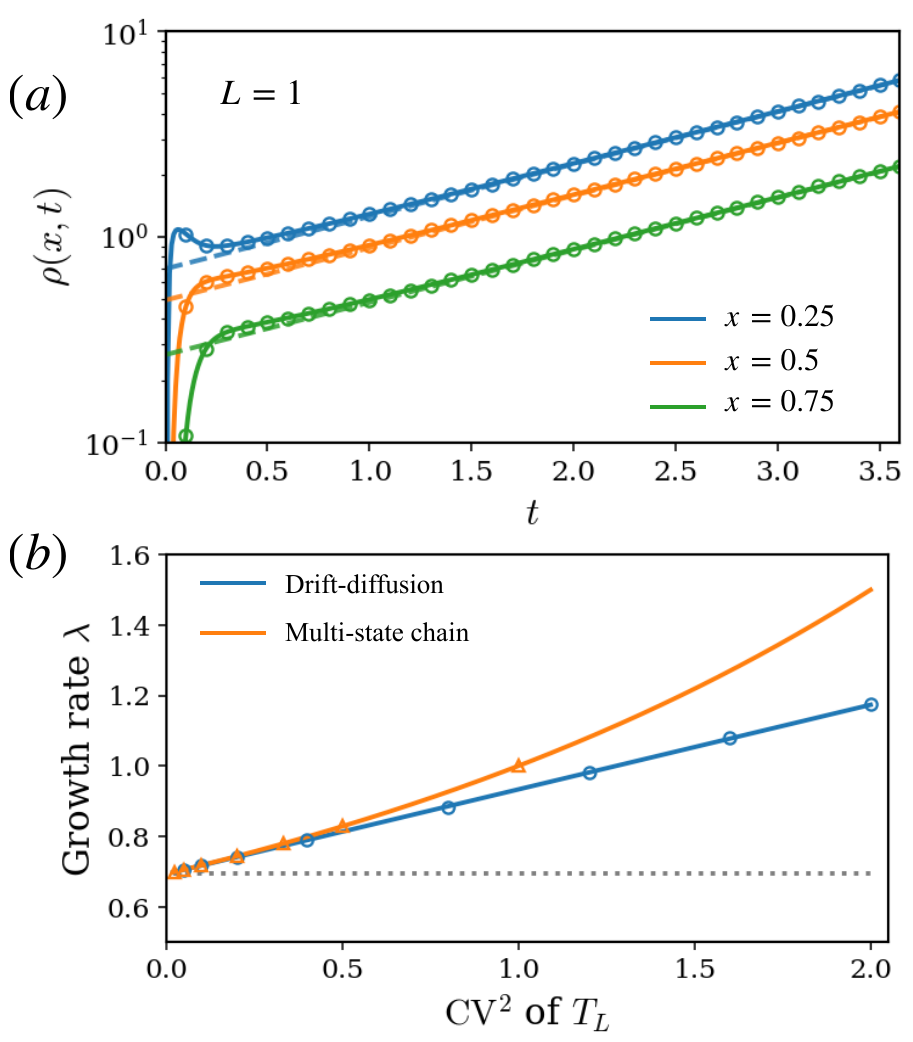}
    \caption{(a) The density $\rho(x,t)$ measured at several positions for a drift-diffusion process with close agreement between the simulations (symbols), exact theory (solid curve) and the asymptotic form (dashed line) in Eq.~\eqref{rhoasymp}. (b) Growth rate $\lambda$ as a function of the first-passage-time variability $\mathrm{CV}^2=\mathrm{Var}(T_L)/\langle T_L\rangle^2$, at fixed mean $\langle T_L\rangle$ and fixed offspring number $m=2$ for representative processes. Increasing CV$^2$ increases $\lambda$. For the discrete multi-state chain, simulation points correspond to Erlang clocks with an integer number $n$ of states, for which $\mathrm{CV}^2=1/n$; hence the markers are not uniformly spaced. Model details and parameters are given in Ref.~\cite{SI}.
}
    \label{fig2}
\end{figure}

To determine the explicit long-time population dynamics, we analyze the behavior of the population density near its dominant pole at $s = \lambda$. Expanding the denominator $1 - m\tilde{f}_L(s)$ around this pole yields the asymptotic population density in the time domain as
\begin{equation}\label{rhoasymp}
    \rho(x,t) \simeq \frac{\tilde{\rho}_0(x,\lambda)}{m\langle T_L e^{-\lambda T_L}\rangle }e^{\lambda t}.
\end{equation}
This set of results, summarized by Eqs.~\eqref{eq:laplacesoln}--\eqref{rhoasymp}, is validated in Fig.~\ref{fig2} through a variety of representative first-passage processes.

\emph{Time-dependent yield.} So far we have treated the replication yield $m$ as a constant parameter. In many threshold-driven replication processes, however, the same internal degree of freedom that sets the replication time also determines the number of descendants produced, so that the yield is actually a function of the first-passage time, or $m\equiv m(T_L)$. Bacteriophage lysis \cite{clokie2011phages}, which is the process by which a bacterium infected by a virus ruptures and releases newly assembled phage, provides a concrete example for such a time-dependent yield  \cite{wang2006lysis}, as depicted in Fig.~\ref{fig:placeholder}(b).

\begin{figure*}
    \centering
    \includegraphics[width=\linewidth]{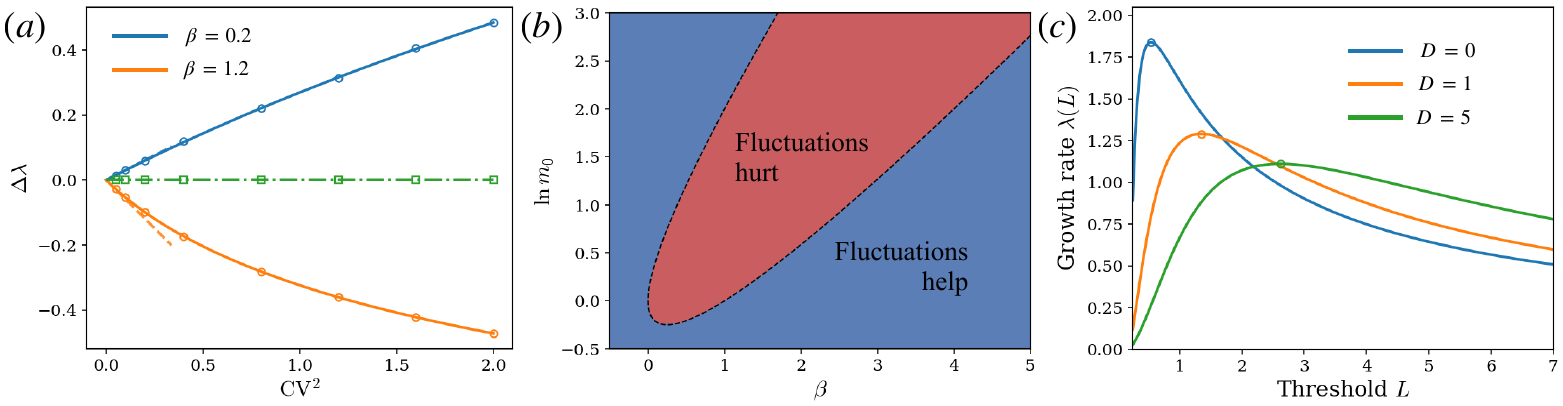}
    \caption{\emph{Time-dependent yield and the yield-delay trade-off.} (a) Change in growth rate, $\Delta\lambda=\lambda-\lambda_{\rm det}$, as a function of the first-passage-time variability $\mathrm{CV}^2$, for drift--diffusion clocks with fixed mean $\langle T_L \rangle = 1$ and fixed yield at the mean, $m(\langle T_L \rangle)=3$. For power-law yields $m(t)=m_0t^\beta$, timing fluctuations can either enhance or suppress growth: they increase $\lambda$ for $\beta=0.2$ but decrease it for $\beta=1.2$. Symbols denote simulations, solid curves denote the exact solution of the generalized Euler--Lotka equation, and the small-fluctuation approximation (dashed line) agrees near the deterministic limit. The exponential-yield control $m(t)=e^{\omega t}$, with $\omega=\ln 3$, gives $\Delta\lambda=0$ for all $\mathrm{CV}^2$ \cite{SI}. (b) Small-noise phase diagram from Eq.~\eqref{eq:criterion_small_noise} for the power-law yield $m(t)=m_0 t^\beta$. Blue indicates parameter values for which small fluctuations enhance growth ($\Delta\lambda>0$), whereas red indicates parameter values for which they reduce growth ($\Delta\lambda<0$). (c) Growth rate $\lambda(L)$ for drift-diffusion with $m(t)=m_0 t$ as obtained in Eq.~\eqref{eq:lambda_linear_LambertW} for different values of $D$, along with the optimal growth rate evaluated at $L_*$ given by Eq.~\eqref{eq:Lstar_linear}. }
    \label{fig3}
\end{figure*}

To account for a time-dependent burst size, $m(t)$, we extend Eq.~\eqref{eq:renewal} as follows,
\begin{equation}
    \rho(x,t) = \rho_0(x,t) + \int_0^t m(t') f_L(t') \cdot \rho(x,t-t') ~dt'.  
\end{equation}
In the long-time limit, the population expands exponentially with a growth rate $\lambda$ that satisfies the generalized version of the Euler-Lotka equation presented in Eq.~\eqref{eq:master},
\begin{equation}\label{eq:master2}
    \langle e^{-\lambda T_L}m(T_L)\rangle \equiv \int_0^\infty e^{-\lambda t} \,m(t) \, f_L(t) \,dt = 1.
\end{equation}
Equation~\eqref{eq:master2} has a useful interpretation: $\lambda$ is the discount rate that makes the expected reproductive value of one first-passage cycle equal to one. It also asserts that the first-passage law tilted by both offspring yield and the exponential cost of delaying descendants, given by
\begin{equation}
    q(t)=e^{-\lambda t}m(t)f_L(t)
\end{equation}
is a normalized probability distribution. By taking the log-ratio of the distributions $f_L(t)$ and $q(t)$ and integrating over all time, we arrive at an exact decomposition for population growth rate as,
\begin{equation}
    \lambda  =  \frac{\langle \ln m(T_L) \rangle + D_{\textrm{KL}}(f_L||q) }{{ \langle T_L \rangle}},
\end{equation}
where $D_{\textrm{KL}}(f_L||q)=\int_0^\infty f_L(t)\ln[f_L(t)/q(t)] ~dt$ is the Kullback-Leibler (KL) divergence \cite{kullback1951information} between the microscopic first-passage distribution and the macroscopic population-level distribution. The first term, $\langle \ln m(T_L)\rangle/\langle T_L\rangle$, is the mean logarithmic yield per replication cycle. The second is a nonnegative correction that quantifies the mismatch between the bare first-passage statistics $f_L(t)$ and the growth-weighted distribution $q(t)$.

To isolate the impact of stochastic fluctuations on the population growth rate, we compare $\lambda$ to the deterministic benchmark $\lambda_\text{det} =\ln\left(m\left(\mu\right)\right)/\mu $, which assumes all replication events occur exactly at the mean first-passage time $\langle T_L\rangle = \mu$.
This form for $\lambda_\mathrm{det}$ naturally follows by using $f_L(t) = \delta(t-\mu)$ in Eq.~\eqref{eq:master2}. For brevity, we define $M(t) = \ln m(t)$. The deviation from the deterministic limit as $\Delta\lambda \equiv \lambda - \lambda_\text{det}$ can thus be expressed as,
\begin{equation}
    \Delta\lambda = \frac{\langle M(T_L) \rangle - M(\mu)}{\mu}+ \frac{D_{\textrm{KL}}(f_L||q)}{\mu}.
\end{equation}
The above exact decomposition can be interpreted in terms of two competing physical contributions: a ``Jensen penalty'' (reflecting the concavity of $M(t)$) which captures the intrinsic cost of yield fluctuations, and the nonnegative KL divergence which quantifies how much $q(t)$ is biased away from $f_L(t)$ by dynamic population-level selection.

\emph{Small fluctuation expansion.} To make the sign of $\Delta\lambda$ explicit, consider weak fluctuations $T_L=\mu+\epsilon$ with $\langle \epsilon\rangle=0$ and $\langle \epsilon^2\rangle=\sigma^2\ll \mu^2$, and write $\lambda=\lambda_{\rm det}+\delta$. Expanding Eq.~\eqref{eq:master2} to leading order gives 
\begin{equation}
\delta= \frac{\sigma^2}{2\mu} \left[\bigl(M'(\mu)-\lambda_{\rm det}\bigr)^2+M''(\mu)
\right]+\mathcal{O}(\sigma^3).
\label{eq:delta_small_noise}
\end{equation}
The expansion is carried out explicitly in Ref.~\cite{SI}. It is evident from Eq.~\eqref{eq:delta_small_noise} that 
\begin{equation}
\bigl(M'(\mu)-\lambda_{\rm det}\bigr)^2+M''(\mu)>0,
\label{eq:criterion_small_noise}
\end{equation}
provides a universal criterion under which small fluctuations enhance population growth. To illustrate this criterion, consider the example $m(t) = m_0 t^\beta$, where $\beta>0$. Equation~\eqref{eq:criterion_small_noise} dictates that small fluctuations in the first-passage process will enhance growth whenever
\begin{equation}
\left(\beta-\ln(m_0\mu^\beta)\right)^2>\beta.
\label{eq:powerlaw_criterion}
\end{equation}
This intricate dependence of the growth rate on timing fluctuations is described in Fig.~\ref{fig3}~(a)~and~(b).

\emph{Yield-delay trade-off and optimization.} When the yield can depend on the first-passage time, a fundamental trade-off arises: waiting longer can potentially increase the eventual yield, but it also delays the onset of all downstream reproduction. Equation~\eqref{eq:master2} therefore defines an optimization problem. For each value of $L$, the first-passage law $f_L(t)$ determines a growth rate $\lambda(L)$, and the optimal threshold $L_*$ is the one that maximizes this growth rate. We first illustrate this trade-off below and then specialize it to bacteriophage lysis.

We consider a drift-diffusion process with the power-law yield $m(t)=m_0t^\beta$. In this case, $\langle T_L^\beta e^{-\lambda T_L}\rangle$ can be computed in closed-form \cite{SI}. For the linear-yield case ($\beta=1$), Eq.~\eqref{eq:master2} admits an exact analytical solution for $\lambda$ in terms of the Lambert $W$ function as follows,
\begin{align}
    \lambda(L) = \frac{D}{L^2}\, W_0\left(\frac{m_0L^2}{2D}\exp\left[\frac{v_{\rm d}L}{2D}\right]\right)^2
    - \frac{v_{\rm d}^2}{4D}.
    \label{eq:lambda_linear_LambertW}
\end{align}
where $D$ is the diffusion constant and $v_{\rm d}$ is the drift. Maximizing Eq.~\eqref{eq:lambda_linear_LambertW} over $L$ gives
\begin{align}\label{eq:Lstar_linear}
 L_* = \frac{e v_{\rm d}+\sqrt{e^2v_{\rm d}^2+8eDm_0}}{2m_0}
\end{align}
and subsequently, $\lambda^* = v_{\rm d}/L_* + D/L_*^2$. In the deterministic limit $D\to0$, this gives $L_*\to ev_{\rm d}/m_0$ and $\lambda^*\to m_0/e$. These results are validated in Fig.~\ref{fig3}(c). Evidently, finite stochasticity shifts the optimum to larger thresholds, but the optimized growth rate reduces with increased fluctuations.

\emph{Application to phage lysis experiments.} A natural application of the time-dependent yield formalism is to bacteriophage lysis. In the language of our formalism, a phage infection cycle separates into two sequential stages. The first is an extracellular adsorption stage: a newly released phage searches for a susceptible host and either adsorbs to the host cell after a random time $A$ or is removed from the population. We denote the adsorption-time density, in the absence of removal, by $g_A(t)$, and the free-phage removal rate by $\delta_P$. The second is an intracellular production stage: after a successful adsorption, the infection proceeds inside the host until an internal lysis variable first reaches a threshold $L$. The corresponding lysis time is $T_L$ and a burst size $m(T_L)$.  Thus the intracellular clock plays two roles simultaneously: it sets the time at which descendants are released and the number of descendants produced.

For a lineage beginning with one newly released phage, the total generation time is $A+T_L$, while the reproductive output is $m(T_L)$. Free-phage loss during the extracellular stage contributes the survival factor $e^{-\delta_P A}$. Assuming that the extracellular adsorption time and intracellular lysis time are independent, the Euler-Lotka condition becomes,
\begin{align}\label{eq:phageEL}
    \tilde g_A(\lambda+\delta_P)
    \left\langle e^{-\lambda T_L}m(T_L)\right\rangle = 1,
\end{align}
where $\tilde g_A(s)=\int_0^\infty e^{-st}g_A(t)~dt$. Equation~\eqref{eq:phageEL} is the phage analogue of Eq.~\eqref{eq:master2}, augmented by an extracellular search factor. 

To connect with the lysis-time measurements of Ref.~\cite{wang2006lysis}, we consider a minimal deterministic description in which intracellular timing fluctuations are neglected, so that $T_L=\tau$. Optimizing the lysis threshold is then equivalent to optimizing the lysis time $\tau$. We use the experimentally measured burst law $m(\tau)=v(\tau-E)$ for $\tau>E$, where $E$ is the eclipse period and $v$ is the maturation rate. By approximating the extracellular stage by a one-step adsorption kernel $g_A(t)= ae^{-at}$, we can see that Eq.~\eqref{eq:phageEL} reduces to
\begin{align}
    \frac{a}{a+\lambda+\delta_P}
    e^{-\lambda \tau}
    v(\tau-E) = 1 .
    \label{eq:phageELdet}
\end{align}
This equation makes the biological trade-off explicit. Increasing $\tau$ allows more intracellular maturation and therefore a larger burst size, but it also delays reproduction. Using $E=28\,\mathrm{min}$, $v\simeq7.7\,\mathrm{phage/min}$, $\delta_P=0.163\,\mathrm{h}^{-1}$, and the independently estimated effective adsorption rate $a\simeq0.178\,\mathrm{h}^{-1}$ from the measurements of Shao and Wang~\cite{shao2008bacteriophage}, the solution to the optimization problem yields $\lambda^*\simeq2.74\,\mathrm{h}^{-1}$ and $\tau^*\simeq49.9\,\mathrm{min}$. These values are close both to Wang's fitted optimum, $(2.71\,\mathrm{h}^{-1},~44.62\,\mathrm{min})$, and to the best measured genotype, $(2.83\pm0.024\,\mathrm{h}^{-1},~ 51.0\pm0.65\,\mathrm{min})$. Together, this analysis serves as a demonstration of the applicability of our framework to real biological processes.

\emph{Discussion.} We introduced a general framework for threshold-driven replication, called \emph{Branching under First-Passage Resetting}, in which replication events are generated endogenously when an internal stochastic state variable first reaches a threshold. At the population level, this construction yields an exact renewal equation that links growth directly to single-trajectory first-passage statistics. If the branching number depends on the first-passage time a fundamental trade-off is revealed, which we discuss in the context of bacteriophage lysis. An important extension of this framework would account for resource constraints on replication, which are present in most biological examples of interest. 

\emph{Acknowledgments.---}AK gratefully acknowledges support from the Complexity Postdoctoral Fellowship of the Santa Fe Institute. JH would like to acknowledge the National Science Foundation Grant Award Number EF--2133863.

\clearpage
\onecolumngrid
\setcounter{page}{1}
\renewcommand{\thepage}{S\arabic{page}}
\setcounter{equation}{0}
\renewcommand{\theequation}{S\arabic{equation}}
\setcounter{figure}{0}
\renewcommand{\thefigure}{S\arabic{figure}}
\setcounter{section}{0}
\renewcommand{\thesection}{S\arabic{section}}
\setcounter{table}{0}
\renewcommand{\thetable}{S\arabic{table}}

\begin{center}
{\large \textbf{Supplemental Material for \emph{Branching under First-Passage Resetting}}}
\end{center}

\onecolumngrid

This Supplemental Material provides details of the stochastic models, parameter choices, asymptotic expansions, and comparisons with experimental bacteriophage data used in the main text. 


\section{Details of models and parameters}
\label{sec:models_parameters}

To illustrate the theory developed in the main text, we use the paradigmatic example of drift-diffusion processes and multi-state chains. In this section, we describe the two processes in detail and discuss the choice of parameters used to prepare Figs.~(2)~and~(3). Throughout, the internal state variable starts at $x(0)=0$, the first-passage threshold is $L$, and the
first-passage time is
\begin{equation}
    T_L=\inf\{t>0:x(t)\geq L\}.
    \label{eq:si-fpt-definition}
\end{equation}
The first-passage density is denoted by $f_L(t)$. For any function $g(t)$, we
write its Laplace transform as
\begin{equation}
    \tilde g(s)=\int_0^\infty e^{-s t}g(t)\,dt .
    \label{eq:si-laplace-definition}
\end{equation}
The fixed offspring number is denoted by $m$, while $m(t)$ denotes a time-dependent offspring yield. The mean first-passage time and squared coefficient of variation are
\begin{align}
    \mu &= \langle T_L\rangle, \\
    \mathrm{CV}^2 &= \frac{\mathrm{Var}(T_L)}{\langle T_L\rangle^2}.
    \label{eq:si-mu-cv-definition}
\end{align}

\subsection{Drift-diffusion process}

The continuous-state model used in Figs.~2(a-b), 3(a), and 3(c) is the classic drift--diffusion process
\begin{equation}
    dx(t)=v_{\rm d}\,dt+\sqrt{2D}\,dW_t
    \label{eq:si-dd-sde}
\end{equation}
where $v_{\rm d}>0$ is the drift velocity, $D\geq 0$ is the diffusion constant,
and $W_t$ is a standard Wiener process. As stated previously, $x(0)=0$. For $D>0$ and $v_{\rm d}>0$, the first-passage density to $L$ is given by
\begin{align}
    f_L(t)
    &=
    \frac{L}{\sqrt{4\pi D t^3}}
    \exp\left[-\frac{(L-v_{\rm d}t)^2}{4Dt}\right].
    \label{eq:si-dd-fpt-density}
\end{align}
In the deterministic limit $D=0$, the first-passage time is $T_L=L/v_{\rm d}$.
The mean, variance, and squared coefficient of variation are
\begin{align}
    \langle T_L\rangle &= \frac{L}{v_{\rm d}}, \\
    \mathrm{Var}(T_L) &= \frac{2DL}{v_{\rm d}^3}, \\
    \mathrm{CV}^2 &= \frac{2D}{Lv_{\rm d}}.
    \label{eq:si-dd-moments}
\end{align}
The Laplace transform of Eq.~\eqref{eq:si-dd-fpt-density} is
\begin{equation}
    \tilde f_L(s)
    =
    \exp\!\left[
    \frac{L}{2D}
    \left(v_{\rm d}-\sqrt{v_{\rm d}^2+4Ds}\right)
    \right].
    \label{eq:si-dd-laplace}
\end{equation}
For fixed offspring number $m$, we can solve the Euler-Lotka equation in closed form, and obtain
\begin{equation}
    \lambda
    =
    \frac{v_{\rm d}}{L}\ln m
    +
    \frac{D}{L^2}(\ln m)^2 .
    \label{eq:si-dd-growth-fixed-yield}
\end{equation}
The deterministic limit $D\to0$ is
\begin{align}
    \lambda = \frac{v_{\rm d}}{L}\ln m .
    \label{eq:si-dd-deterministic-limit}
\end{align}
Interestingly, we see that, at fixed $\langle T_L\rangle$, the growth rate increases monotonically with $D$, or equivalently with $\mathrm{CV}^2$.

\subsection{Multi-state chain (Erlang and Gamma clocks)}

The discrete-state model used in Fig.~2(b) is a unidirectional multi-state chain. In this case, the internal state variable takes values in the nonnegative integers, and the first-passage threshold $L$ is a positive integer. The process begins in state $0$ and advances irreversibly through $L$ exponential stages, $0 \to 1 \to \cdots \to L$, with transition rate $r$ between consecutive states. The first-passage time $T_L$ is the time at which state $L$ is first reached. Equivalently,
\begin{align}
    T_L
    &=
    \sum_{i=1}^{L} E_i ,
    \label{eq:si-erlang-sum}
\end{align}
where the $E_i$ are independent exponential random variables with rate $r$. Thus $T_L$ has the Erlang density
\begin{align}
    f_L(t)
    &=
    \frac{r^L t^{L-1} e^{-rt}}{(L-1)!},
    \qquad t>0 .
    \label{eq:si-erlang-density}
\end{align}
To compare clocks at fixed mean $\mu=\langle T_L\rangle$, we choose
\begin{align}
    r
    &=
    \frac{L}{\mu}.
    \label{eq:si-erlang-rate}
\end{align}
With this convention,
\begin{align}
    \nonumber\langle T_L\rangle    &=
    \mu,\\
    \mathrm{Var}(T_L)
    &=
    \frac{\mu^2}{L},\\
    \nonumber\mathrm{CV}^2
    &=
    \frac{1}{L},
    \label{eq:si-erlang-moments}
\end{align}
and its Laplace transform is
\begin{align}
    \tilde{f}_L(s)
    &=
    \left(1+\frac{\mu s}{L}\right)^{-L}.
\end{align}
For fixed offspring number $m$, the Euler--Lotka equation $m\tilde{f}_L(\lambda)=1$ can therefore be solved in closed form, giving
\begin{align}
    \lambda(L)
    &=
    \frac{L}{\mu}
    \left(m^{1/L}-1\right).
    \label{eq:si-erlang-growth}
\end{align}
In the limit $L\to\infty$, the Erlang clock converges to a deterministic clock with $T_L=\mu$, and Eq.~\eqref{eq:si-erlang-growth} reduces to
\begin{align}
    \lambda_{\rm det}
    &=
    \frac{\ln m}{\mu}.
    \label{eq:si-erlang-det-limit}
\end{align}

For Fig.~2(b), we also use the continuous Gamma-clock extension of this Erlang family. We write the Gamma density as
\begin{align}
    f_{\Gamma}(t;k,\theta)
    &=
    \frac{t^{k-1}e^{-t/\theta}}{\Gamma(k)\theta^k},
    \label{eq:si-gamma-density}
\end{align}
where $k$ is the shape and $\theta$ is the scale. Fixing the mean and squared coefficient of variation gives
\begin{align}
    k
    &=
    \frac{1}{\mathrm{CV}^2},
    &
    \theta
    &=
    \frac{\mu}{k}
    =
    \mu\,\mathrm{CV}^2 .
    \label{eq:si-gamma-params}
\end{align}
The Laplace transform is
\begin{align}
    \tilde{f}_{\Gamma}(s)
    &=
    (1+\theta s)^{-k}
    =
    \left(1+\mu\,\mathrm{CV}^2 s\right)^{-1/\mathrm{CV}^2}.
    \label{eq:si-gamma-laplace}
\end{align}
Substituting Eq.~\eqref{eq:si-gamma-laplace} into $m\tilde{f}_{\Gamma}(\lambda)=1$ gives
\begin{align}
    \lambda_{\Gamma}
    &=
    \frac{1}{\mu\,\mathrm{CV}^2}
    \left(m^{\mathrm{CV}^2}-1\right).
    \label{eq:si-gamma-growth}
\end{align}
When $k=L$ is a positive integer, Eq.~\eqref{eq:si-gamma-growth} reduces exactly to the Erlang multi-state-chain result in Eq.~\eqref{eq:si-erlang-growth}, with $\mathrm{CV}^2=1/L$. Thus the Gamma curve in Fig.~2(b) is a continuous interpolation through the Erlang clocks. The literal multi-state-chain interpretation applies at integer $L$, while noninteger $k$ should be viewed as a Gamma-clock continuation.

\subsection{Numerical parameters used in the figures}

\textbf{Figure 2(a).} We use a drift-diffusion process with
\begin{align*}
    L &= 1,\\
    v_{\rm d} &= 0.5,\\
    D &= 0.5,\\
    m &= 2 .
\end{align*}
The density $\rho(x,t)$ was evaluated at
\begin{align*}
    x/L &= 0.25,~0.5,~0.75.
\end{align*}

\textbf{Figure 2(b).} For Fig.~2(b), all clocks were compared at fixed mean first-passage time and fixed offspring number,
\begin{align*}
    \mu &= 1,\\
    m &= 2 .
\end{align*}
For the drift--diffusion clock, we used
\begin{align*}
    L &= 1,\\
    v_{\rm d} &= 1,\\
    D &= \frac{\mathrm{CV}^2}{2}.
\end{align*}
This choice fixes $\langle T_L\rangle=L/v_{\rm d}=1$. The deterministic reference line is
\begin{align*}
    \lambda_{\rm det} &= \ln 2 .
\end{align*}

For the Erlang multi-state chain in Fig.~2(b), the threshold $L$ is the integer number of exponential stages. The simulation points were generated at
\begin{align*}
    L = 40,~20,~10,~5,~3,~2,~1,
\end{align*}
with transition rate
\begin{align*}
    r &= \frac{L}{\mu}.
\end{align*}
The smooth Gamma-clock curve was evaluated using
\begin{align*}
    k &= \frac{1}{\mathrm{CV}^2},\\
    \theta &= \frac{\mu}{k}
            =
            \mu \mathrm{CV}^2 .
\end{align*}
For $\mu=1$ and $m=2$, Eq.~\eqref{eq:si-gamma-growth} becomes
\begin{align*}
    \lambda_{\Gamma}
    &=
    \frac{2^{\mathrm{CV}^2}-1}{\mathrm{CV}^2}.
\end{align*}
The deterministic value at $\mathrm{CV}^2=0$ is understood as the limiting value $\lambda_{\Gamma}\to\ln 2$.

\textbf{Figure 3(a).} We used drift-diffusion process times with
\begin{align*}
    \mu &= 1,\\
    L &= 1,\\
    v_{\rm d} &= 1,\\
    D &= \frac{\mathrm{CV}^2}{2}.
\end{align*}
The time-dependent yield was
\begin{align*}
    m(t) &= m_0 t^\beta .
\end{align*}
We fixed the yield at the mean first-passage time by setting $m(\mu) = 3$. Since $\mu=1$, this gives $m_0=3$. The two power-law exponents shown were $\beta = 0.2~$and$~1.2$. The deterministic benchmark used in $\Delta\lambda=\lambda-\lambda_{\rm det}$ is
\begin{align*}
    \lambda_{\rm det}
    &=
    \ln 3 .
\end{align*}
The horizontal reference curve corresponds to the exponential yield
\begin{align*}
    m(t) &= \exp(\omega t),\\
    \omega &= \ln 3 .
\end{align*}
For this choice, $\Delta\lambda=0$ for all first-passage-time distributions with positive support. This can be seen easily from the generalized Euler-Lotka equation, where upon substituting $m(t)=e^{\omega t}$, we get
\begin{align*}
    \left\langle e^{-(\lambda-\omega)T_L} \right\rangle = 1 .
\end{align*}
Equivalently,
\begin{align*}
    \tilde f_L(\lambda-\omega)=1,
\end{align*}
Since $T_L>0$, we have $\tilde f_L(0)=1$, while $\tilde f_L(s)<1$ for $s>0$. Thus the unique solution is
\begin{align*}
    \lambda-\omega=0,
    \qquad \text{so} \qquad
    \lambda=\omega .
\end{align*}
The deterministic benchmark at the same mean first-passage time $\mu=\langle T_L\rangle$ is
\begin{align*}
    \lambda_{\rm det}
    = \frac{\ln m(\mu)}{\mu} = \frac{\ln e^{\omega \mu}}{\mu} = \omega .
\end{align*}
Therefore,
\begin{align*}
    \Delta \lambda = \lambda-\lambda_{\rm det} = 0.
\end{align*}
Thus the exponential yield is exactly fluctuation-neutral: changing the variability of $T_L$ does not change the growth rate relative to the deterministic clock with the same mean.

\textbf{Figure 3(b).} The phase diagram was computed from the small-fluctuation criterion
\begin{align*}
    \left[\beta-\ln(m_0\mu^\beta)\right]^2-\beta > 0,
\end{align*}
with $\mu=1$. Blue indicates parameter values for which weak timing fluctuations enhance growth, while red indicates parameter values for which they reduce growth.

\textbf{Figure 3(c).} For Fig.~3(c), we used the linear yield
\begin{align*}
    m(t) &= m_0 t,
\end{align*}
with
\begin{align*}
    \beta &= 1,\\
    m_0 &= 5,\\
    v_{\rm d} &= 1 .
\end{align*}
The three curves correspond to $D=0,1,~$and$~5$. The optimal thresholds and growth rates for the displayed parameters are
\begin{align*}
    D=0:\qquad
    L_* &\simeq 0.544, \qquad
    \lambda_* \simeq 1.839,\\
    D=1:\qquad
    L_* &\simeq 1.349,
    \qquad
    \lambda_* \simeq 1.290,\\
    D=5:\qquad
    L_* &\simeq 2.619,
    \qquad
    \lambda_* \simeq 1.111 .
\end{align*}

\section{Small fluctuation expansion}
\label{sec:small_fluctuation_expansion}

In this section, we carry out the derivation of Eq.~(12)~and~(13) from the main text. Our goal is to expand the generalized Euler-Lotka equation corresponding to time-dependent yield given by
\begin{equation}
    \left\langle  e^{-\lambda T_L}m(T_L)\right\rangle=1
    \label{eq:SI_small_noise_EL_start}
\end{equation}
around the deterministic limit.  Write
\begin{equation}
    T_L=\mu+\varepsilon,
    \qquad
    \langle\varepsilon\rangle=0,
    \qquad
    \langle\varepsilon^2\rangle=\sigma^2\ll\mu^2,
    \label{eq:SI_noise_definitions}
\end{equation}
and assume that higher central moments scale regularly, so that $\langle |\varepsilon|^3\rangle=O(\sigma^3)$.  We set
\begin{equation}
    \lambda=\lambda_{\rm det}+\delta,
    \qquad
    \lambda_{\rm det}=\frac{M(\mu)}{\mu}.
\end{equation}
Note that we are using the definition $M(t) = \ln m(t)$. By rewriting Eq.~\eqref{eq:SI_small_noise_EL_start} as $\left\langle  e^{M(T_L) -\lambda T_L}\right\rangle=1$, we can expand the exponent as follows,
\begin{align}
    M(\mu+\varepsilon)-(\mu+\varepsilon)\lambda
    &=M(\mu)-\lambda_{\rm det}\mu
    +\left[M'(\mu)-\lambda_{\rm det}\right]\varepsilon
    +\frac{1}{2}M''(\mu)\varepsilon^2
    -\delta\mu
    -\delta\varepsilon
    +O(\varepsilon^3).
    \label{eq:SI_exponent_expansion}
\end{align}
The first term vanishes by definition of $\lambda_{\rm det}$.  Since $\delta=O(\sigma^2)$, the term $\delta\varepsilon$ contributes only at $O(\sigma^3)$ after averaging.  Expanding the exponential and keeping all terms of order $\sigma^2$ gives
\begin{align}
    1
    &=\left\langle
    \exp\left[
    \left(M'(\mu)-\lambda_{\rm det}\right)\varepsilon
    +\frac{1}{2}M''(\mu)\varepsilon^2
    -\delta\mu
    +O(\sigma^3)
    \right]
    \right\rangle
    \\
    &=1-\delta\mu
    +\frac{\sigma^2}{2}M''(\mu)
    +\frac{\sigma^2}{2}\left(M'(\mu)-\lambda_{\rm det}\right)^2
    +O(\sigma^3).
    \label{eq:SI_exp_average}
\end{align}
Thus
\begin{equation}
    \delta=\frac{\sigma^2}{2\mu}
    \left[
    \left(M'(\mu)-\lambda_{\rm det}\right)^2+M''(\mu)
    \right]
    +O(\sigma^3).
    \label{eq:SI_small_noise_result}
\end{equation}
Weak timing fluctuations enhance growth when
\begin{equation}
    \left(M'(\mu)-\lambda_{\rm det}\right)^2+M''(\mu)>0.
    \label{eq:SI_small_noise_criterion}
\end{equation}
The two terms in Eq.~\eqref{eq:SI_small_noise_criterion} have distinct interpretations.  The squared term is always positive and is the local version of the population-level bias encoded by $D_{\rm KL}(f_L || q)$.  The curvature term is the local Jensen contribution; it is negative for concave $M(t)$ and positive for convex $M(t)$.

For the power-law yield
\begin{equation}
    m(t)=m_0t^\beta,
    \qquad
    \beta>0,
\end{equation}
we have
\begin{equation}
    M(t)=\ln m_0+\beta\ln t,
    \qquad
    M'(\mu)=\frac{\beta}{\mu},
    \qquad
    M''(\mu)=-\frac{\beta}{\mu^2},
\end{equation}
and
\begin{equation}
    \lambda_{\rm det}=\frac{\ln(m_0\mu^\beta)}{\mu}.
\end{equation}
Equation~\eqref{eq:SI_small_noise_result} becomes
\begin{equation}
    \delta=\frac{\sigma^2}{2\mu^3}
    \left[
    \left(\beta-\ln(m_0\mu^\beta)\right)^2-\beta
    \right]
    +O(\sigma^3).
    \label{eq:SI_powerlaw_small_noise_delta}
\end{equation}
Thus small fluctuations enhance growth when
\begin{equation}
    \left(\beta-\ln(m_0\mu^\beta)\right)^2>\beta,
    \label{eq:SI_powerlaw_small_noise_criterion}
\end{equation}
which is the criterion plotted in Fig.~3(b).  The red region in that figure is centered near the curve $\ln(m_0\mu^\beta)=\beta$, where the logarithmic yield slope is matched to the deterministic growth rate and the negative curvature of $\ln m(t)$ dominates the fluctuation correction.

\section{Optimization}
\label{sec:optimization}

Once the offspring yield depends on the first-passage time, the threshold $L$ is not merely a time-setting parameter.  Increasing $L$ typically delays reproduction, but it can also increase the realized yield by allowing the internal production process to run longer. This creates a trade-off which was discussed in the main text. Here, we go over the calculations and derivations which led to Eq.~(15)~and~(16) in the main text.

The growth rate is determined implicitly by the generalized Euler-Lotka equation in Eq.~\eqref{eq:SI_small_noise_EL_start}. For drift-diffusion and power-law yield $m(t)=m_0t^\beta$, the required moment-Laplace transform can be evaluated analytically.  In particular, using 
\begin{equation}
    \int_0^\infty t^{\nu-1}\exp(-At-B/t)\,dt
    =2\left(\frac{B}{A}\right)^{\nu/2}K_\nu(2\sqrt{AB}),
\end{equation}
where $K_\nu$ is the modified Bessel function of the second kind, we get
\begin{equation}
\left\langle T_L^\beta  e^{-\lambda T_L}\right\rangle
=
\frac{L e^{v_{\rm d}L/(2D)}}{\sqrt{\pi D}}
\left(\frac{L^2}{v_{\rm d}^2+4D\lambda}\right)^{(\beta-1/2)/2}
K_{\beta-1/2}\left(\frac{L}{2D}\sqrt{v_{\rm d}^2+4D\lambda}\right).
\label{eq:SI_momentLaplace_Bessel}
\end{equation}
The growth rate for power-law yield is consequently determined by
\begin{equation}
\frac{m_0L e^{v_{\rm d}L/(2D)}}{\sqrt{\pi D}}
\left(\frac{L^2}{v_{\rm d}^2+4D\lambda}\right)^{(\beta-1/2)/2}
K_{\beta-1/2}\left(\frac{L}{2D}\sqrt{v_{\rm d}^2+4D\lambda}\right)=1.
\label{eq:SI_EL_powerlaw_Bessel}
\end{equation}
For arbitrary $\beta$, Eq.~\eqref{eq:SI_EL_powerlaw_Bessel} must be solved numerically. However, the linear-yield case $m(t)=m_0t$ is special, not only because of its applicability to bacteriophage lysis, but also because it admits a closed form solution.  Using
\begin{equation}
    \left\langle T_L e^{-\lambda T_L}\right\rangle
    =-\frac{d}{d\lambda}\tilde f_L(\lambda)
    =\frac{L}{\sqrt{v_{\rm d}^2+4D\lambda}}
    \exp\left[\frac{L}{2D}\left(v_{\rm d}-\sqrt{v_{\rm d}^2+4D\lambda}\right)\right],
\end{equation}
Eq.~\eqref{eq:SI_small_noise_EL_start} becomes
\begin{equation}
    m_0L\frac{\exp\left[\frac{L}{2D}\left(v_{\rm d}-\sqrt{v_{\rm d}^2+4D\lambda}\right)\right]}{\sqrt{v_{\rm d}^2+4D\lambda}}=1.
    \label{eq:SI_linear_yield_EL_intermediate}
\end{equation}
Let
\begin{equation}
    R=\sqrt{v_{\rm d}^2+4D\lambda},
    \qquad
    y=\frac{RL}{2D}.
\end{equation}
Equation~\eqref{eq:SI_linear_yield_EL_intermediate} can be written as
\begin{equation}
    y e^y=\frac{m_0L^2}{2D}\exp\left(\frac{v_{\rm d}L}{2D}\right).
\end{equation}
Thus, on the principal branch of the Lambert function,
\begin{equation}
    y=W_0\left(\frac{m_0L^2}{2D}\exp\left[\frac{v_{\rm d}L}{2D}\right]\right),
\end{equation}
and
\begin{equation}
    \lambda(L)=\frac{D}{L^2}
    W_0\left(\frac{m_0L^2}{2D}\exp\left[\frac{v_{\rm d}L}{2D}\right]\right)^2
    -\frac{v_{\rm d}^2}{4D}.
    \label{eq:SI_linear_yield_lambda_L}
\end{equation}
This is Eq.~(15) of the Letter. The optimum can also be found analytically.  Differentiating Eq.~\eqref{eq:SI_linear_yield_lambda_L} with respect to $L$ and using $W_0'(z)=W_0(z)/[z(1+W_0(z))]$ gives the stationarity condition
\begin{equation}
    W_0\left(\frac{m_0L^2}{2D}\exp\left[\frac{v_{\rm d}L}{2D}\right]\right)=1+\frac{v_{\rm d}L}{2D}.
    \label{eq:SI_W_stationarity}
\end{equation}
Substitution into the defining equation for $W_0$ yields
\begin{equation}
    m_0L^2- e v_{\rm d}L-2 e D=0.
\end{equation}
The positive solution is
\begin{equation}
    L^*=\frac{ e v_{\rm d}+\sqrt{ e^2v_{\rm d}^2+8 e Dm_0}}{2m_0}.
    \label{eq:SI_linear_yield_Lstar}
\end{equation}
At this optimum, Eq.~\eqref{eq:SI_W_stationarity} simplifies the growth rate to
\begin{equation}
    \lambda^*=\frac{v_{\rm d}}{L^*}+\frac{D}{(L^*)^2}.
    \label{eq:SI_linear_yield_lambdastar}
\end{equation}
In the deterministic limit $D\to0$, Eqs.~\eqref{eq:SI_linear_yield_Lstar} and \eqref{eq:SI_linear_yield_lambdastar} reduce to
\begin{equation}
    L^*\to\frac{ e v_{\rm d}}{m_0},
    \qquad
    \lambda^*\to\frac{m_0}{e}.
\end{equation}

\section{Application to Bacteriophage Lysis}
\label{sec:viral_lysis}

We now show how the \emph{Branching under First-Passage Resetting} (BFPR) framework applies to phage lysis data. It is useful to write the renewal cycle from the perspective of a newly released phage. We decompose an infection cycle into two stages. 
\begin{enumerate}
    \item First, a free phage searches for a susceptible host cell. We denote by $A$ the time needed to find and infect a host cell (adsorption time) and by $g_A(t)$ the adsorption-time density in the absence of free-phage removal. Free phage are removed at rate $\delta_P$, so $g_A(t)\mathrm{e}^{-\delta_P t}$ is the effective density for adsorption at time $t$ before removal.
    \item Next, after a phage successfully infects a host, intracellular assembly proceeds until a lysis threshold $L$ is reached at first-passage time $T_{L}$. This initiates lysis and is referred to as the lysis time. 
\end{enumerate}
Consequently, the total time for one cycle to be completed (instance of renewal) is $A + T_L$. As a simplifying assumption, we neglect stochasticity in the intracellular lysis time and set $T_{L}=\tau$. Then the viral extension of the generalized Euler--Lotka relation is given by
\begin{equation}
\tilde g_A(\lambda+\delta_P) e^{-\lambda \tau}m(\tau)=1.
\label{eq:viral_EL}
\end{equation}
Equation~(\ref{eq:viral_EL}) is the natural BFPR description of viral growth: the intracellular first-passage problem enters through $\tau$ and the time-dependent yield $m(t)$, while the extracellular search stage enters through the Laplace transform of $g_A$. 

\noindent To make direct contact with the data, we refer to Wang 2006. Wang showed that the burst size is approximately linear in lysis time, 
\begin{equation}
m(\tau)=
\begin{cases}
0, & \tau\le E,\\[3pt]
v(\tau-E), & \tau>E.
\end{cases}
\label{eq:burst_linear}
\end{equation}
where the maturation rate $v\simeq 7.7$ phage/min and eclipse period $E\simeq 28$ min. Furthermore, Wang notes that $\delta_P = 0.163\,\mathrm{h^{-1}}$. For the extracellular stage we adopt a minimal one-step search kernel
\begin{equation}
g_A(t)=a e^{-at},
\qquad
\tilde g_A(s)=\frac{a}{a+s},
\label{eq:one_step_search}
\end{equation}
where $a$ is the effective rate of successful reinfection after lysis. Consequently, Eq.~\eqref{eq:viral_EL} reduces to
\begin{equation}
\frac{a}{a+\lambda+\delta_P}\,e^{-\lambda\tau}\,v(\tau-E)=1,
\qquad \tau>E.
\label{eq:deterministic_viral_EL}
\end{equation}
Here and below, all times entering exponentials or rates are understood to be expressed in the same units. Along the solution branch $\lambda(\tau)$, the optimum is defined by $d\lambda/d\tau=0$. Differentiating the logarithm of Eq.~\eqref{eq:deterministic_viral_EL} therefore gives
\begin{equation}
0=-\lambda^*+\frac{1}{\tau^*-E},
\end{equation}
so that
\begin{equation}
\tau^*=E+\frac{1}{\lambda^*}.
\label{eq:tau_star_viral}
\end{equation}
Substituting Eq.~\eqref{eq:tau_star_viral} back into Eq.~\eqref{eq:deterministic_viral_EL} yields the scalar equation for the optimal growth rate,
\begin{equation}
\frac{a}{a+\lambda^*+\delta_P}\,\frac{v}{e\lambda^*}\,e^{-\lambda^* E}=1.
\label{eq:lambda_star_viral}
\end{equation}
Equations~(\ref{eq:tau_star_viral}) and (\ref{eq:lambda_star_viral}) are the deterministic BFPR benchmark that we compare with experiment.

The key modeling issue is how to set the extracellular search rate $a$. We use the independent adsorption measurements of Shao and Wang. They found an adsorption coefficient $k_{\rm HA}=9.90\times10^{-9}\,\mathrm{mL\,cell^{-1}\,min^{-1}}$ for the high-adsorption $\mathrm{stf}^{+}$ background. Evaluated at Wang's host density, this gives
\begin{equation}
a\equiv a_{\rm eff}\approx k_{\rm HA}N_0\times 60
=(9.90\times10^{-9})(3\times10^5)\times 60
=0.178\,\mathrm{h^{-1}}.
\label{eq:a_eff}
\end{equation}
We emphasize that Eq.~\eqref{eq:a_eff} is used only as an externally validated estimate of the effective search scale in Wang's assay. 

Using $E=28/60\,\mathrm{h}$, $v=7.7\times 60=462\,\mathrm{h^{-1}}$, $\delta_P=0.163\,\mathrm{h^{-1}}$, and $a=0.178\,\mathrm{h^{-1}}$, Eq.~\eqref{eq:lambda_star_viral} becomes
\begin{equation}
\frac{0.178}{0.178+\lambda^*+0.163}\,\frac{462}{e\lambda^*}\,e^{-0.4667\lambda^*}=1,
\end{equation}
which gives
\begin{equation}
\lambda^*=2.74\,\mathrm{h^{-1}}.
\end{equation}
Equation~(\ref{eq:tau_star_viral}) then yields
\begin{equation}
\tau^*=28\,\mathrm{min}+\frac{60}{2.74}=49.9\,\mathrm{min}.
\end{equation}
These predictions compare well with Wang's quadratic optimum,
\begin{equation}
(\tau_{\rm opt}^{\rm fit},w_{\rm max}^{\rm fit})=(44.62\,\mathrm{min},2.71\,\mathrm{h^{-1}}),
\end{equation}
and with the best directly measured genotype in Wang's Table~2, wild-type $S$, which has
\begin{equation}
(\tau_{\rm wt},w_{\rm wt})=(51.0\pm0.65\,\mathrm{min},2.83\pm0.024\,\mathrm{h^{-1}}).
\end{equation}
Thus the deterministic BFPR benchmark, with intracellular lysis treated as deterministic and the extracellular stage represented by a one-step effective search kernel whose scale is fixed independently from Shao--Wang, predicts both the optimal lysis time and the corresponding growth rate of the Wang dataset to good accuracy.

\end{document}